# Deep Residual Shrinkage Networks for EMG-based Gesture Identification

Yueying Ma[1], Chengbo Wang[2], Chengenze Jiang[3], Zimo Li[4]

School of Physical Science and Technology, Lanzhou University, Lanzhou, Gansu 730107, China
School of Advanced Technology, Xi'an Jiaotong-Liverpool University, Suzhou, Jiangsu 215123, China
Chongqing Jiangbei Bachuan Quantum Middle School, Chongqing 400000, China
School of Chengdu Shishi Cambridge A-Level Centre, Chengdu, Sichuan 610041, China

*Abstract*—This work introduces a method for high-accuracy EMG based gesture identification. A newly developed deep learning method, namely, deep residual shrinkage network is applied to perform gesture identification. Based on the feature of EMG signal resulting from gestures, optimizations are made to improve the identification accuracy. Finally, three different algorithms are applied to compare the accuracy of EMG signal recognition with that of DRSN. The result shows that DRSN excel traditional neural networks in terms of EMG recognition accuracy. This paper provides a reliable way to classify EMG signals, as well as exploring possible applications of DRSN.

## I. Introduction

Electrography (EMG) is a diagnostic procedure to assess the health of muscles and the nerve cells that control motor neurons. Surface electromyography (sEMG) is scanning electromyography on surface, and it is non-invasive . sEMG can nowadays theoretically be used to diagnose neuromuscular disease and low back pain and assess the prognosis of diseases involving muscle lesions. However, the clinical efficacy of sEMG has not been proven in peer-reviewed medical literature. In the future, sEMG might be used to monitor the progression of neurological and muscle diseases and monitor dental grinding.

EMG signals are widely occurred in testing muscle or neuron activities and are becoming increasingly important in clinical, biomedical, prosthetic repair devices, human-computer interaction, and more. EMG signals have following features. Firstly, they can be detected in wide are of the muscle, which makes them hard to be located and detected. Secondly, the frequency band of them is relatively narrow, usually from 20Hz to 500Hz . Thirdly, they have a low-signal resolution leading to inaccuracy. The last one is the highly susceptible to movement artifact . There are many kinds of methods to analyze the signals. One of them is neuro-fuzzy matrix modifier [1]. It calculates user's hand force vector based on the estimated joint torque to obtain hand acceleration to estimate the user's hand trajectory. This method could identity different postures made by different users effectively. Another method proposed by J. Bird et.al [2] is cross domain multilayer perceptron, which can be applied in both EMG gesture classification and EEG mental state recognition. His method shows that knowledge transfer is possible even without training being required. The third one is CNN Transfer Learning [2,3]. It represents biological waves as images and then hyperparameters are derived. CNN transfer learning approach is only successful in the case of transformation from EMG to EEG but not vice-versa [2].

In general, there are two existing categories of algorithms to process sEMG signals, signal analysis-based methods and machine-learning-powered methods. For signal analysis-based methods, they can be mainly divided by three parts: time domain analysis, frequency domain analysis, spatial information. These traditional methods have several disadvantages. They are sensitive to noise and non-stationarities as well as they have no temporal dynamics and reduced spectral precision.

For machine learning, there are such models, Logistic Regression, Random Forest, Convolutional Neural Networks (CNN), Deep Residual Networks (DRN) and Deep Residual Shrinkage Networks (DRSN). The machine learning has lower calculation cost. And it is also easy to understand and implement making it more practical in the society. When testing the data set, machine learning runs faster and scales well to large databases. For outcomes, it can visualize the analysis and it has strong robustness and fault tolerance to noise data. However, for traditional deep learning methods, it is often a difficult problem for parameter optimization. The back layer must be propagated by gradient of the error function by layer and the gradient of the error function gradually gets more and more inaccurate after many layers. Therefore, it is ineffective to optimize the trainable parameters of the initial layer (that is, the layer close to the input layer).

Convolutional Neural Networks (CNN) is a type of artificial neural network used in imagerecognition and processing that is specificallydesigned to process pixel data. For DRN, it is an attractive variant of CNN eases parameter optimization by using identity shortcuts.  In DRN, not only does the gradient propagate back layer by layer, but also flows directly back to the original layer by identifying shortcuts. Whereas it exists the disadvantages. When processing high noise vibration signals, the DRN's feature learning ability tends to be reduced. The convolution kernel used in DRN as a local feature extractor may not be able to detect fault related features due to noise interference. From this, it is often not enough to identify the fault correctly by using the advanced features learned at the output layer are. Therefore, a new way called DRSN[4] was developed.

In this paper, two deep residual shrinkage networks (DRSNs), namely the deep Residual shrinkage network with channel sharing threshold (DRSN-CS) and the deep residual shrinkage network with channel sharing threshold (DRSN-CW), are developed to improve the ability of DRN to learn features from high-noise vibration signals, and finally achieve the goal of high diagnostic accuracy. The main contributions are summarized below.

1) To effectively eliminate the noise correlation characteristics it inserts the soft threshold (a popular contraction function) into the deep structure acting as a nonlinear transform layers.

2) A specially designed subnetwork is used to determine the threshold adaptive. As results, each vibration signal has its own set of thresholds.

3) The soft threshold takes two kinds of thresholds into account, the channel-shared threshold and the channel-wise threshold, which are where the terms DRSN-CS and DRSN-CW come from [7].

## II. THEORY OF DRSN

As stated in Section I, DRSN has broad application prospects in the field of high-noise data recognition, Therefore, this section mainly discusses the reasons for proposing DRSN, indicating that soft thresholding is the core step of many signal noise reduction algorithms, introducing the attention mechanism and soft thresholding under the deep attention mechanism, and predicting that the deep residual shrinkage network may have Broader versatility.

### A. Proposal of Deep Residual Shrinkage Network

The deep residual shrinkage network is an improved version of the deep residual network, which is an integration of the deep residual network, attention mechanism and soft threshold function. To a certain extent, the working principle of the deep residual shrinkage network can be understood as: paying attention to unimportant features through the attention mechanism and setting them to zero through the soft threshold function; in other words, paying attention to important features through the attention mechanism Features, and keep them, thereby enhancing the ability of deep neural networks to extract useful features from noisy signals.

For the main purpose to develop deep residual shrinkage networks, first, when classifying samples, there will inevitably be some noise in the sample, such as Gaussian noise, pink noise, Laplacian noise, etc. More broadly speaking, the sample is likely to contain information that has nothing to do with the current classification task, and this information can also be understood as noise. These noises may adversely affect the classification effect. (Soft thresholding is a key step in many signal noise reduction algorithms). Secondly, even in the same sample set, the amount of noise in each sample is often different.

This is similar to the attention mechanism; taking an image sample set as an example, the location of the target object in each picture may be different; the attention mechanism can pay attention to the location of the target object for each picture.

### B. Soft Thresholding

Soft thresholding is the core step of many signal noise reduction algorithms. Features whose absolute value is less than a certain threshold is deleted, and features whose absolute value is greater than this threshold is shrunk toward zero. It can be achieved by the following formula:

$$y = \begin{cases} x - \tau & x > \tau \\ 0 & -\tau \leq x \leq \tau \\ x + \tau & x < -\tau \end{cases} \quad (1)$$

The derivative of the soft threshold output with respect to the input is

$$\frac{\partial y}{\partial x} = \begin{cases} 1 & x > \tau \\ 0 & -\tau \leq x \leq \tau \\ 1 & x < -\tau \end{cases} \quad (2)$$

It can be seen from equation (2) that the derivative of soft thresholding is either 1 or 0. This property is the same as the ReLU activation function. Therefore, soft thresholding can also reduce the risk of gradient dispersion and gradient explosion for deep learning algorithms.

In the soft thresholding function, the threshold setting must meet two conditions: First, the threshold is a positive number; second, the threshold cannot be greater than the maximum value of the input signal, otherwise the output will be all zero. At the same time, the threshold should also meet the third condition: each sample should have its own independent threshold according to its own noise content. This is because the noise content of many samples is often different. For example, it is often the case that in the same sample set, sample A contains less noise, and sample B contains more noise. Then, if the soft threshold is used in the noise reduction algorithm, sample A should use a larger threshold, and sample B should use a smaller threshold. In deep neural networks, although these features and thresholds lose their clear physical meaning, the basic principles are still the same. In other words, each sample should have its own independent threshold based on its own noise content.

### C. Attention Mechanism

The attention mechanism is relatively easy to understand in the field of computer vision. The animal's visual system can quickly scan the entire area, find the target object, and then focus on the target object to extract more details while suppressing irrelevant information.

Squeeze-and-Excitation Network (SENet) is a newer deep learning method under the attention mechanism. In different samples, different feature channels have different contributions in classification tasks. SENet uses a small sub-network as shown in Fig. 1 to obtain a set of weights, and then multiplies this set of weights with the characteristics of each channel to adjust the size of each channel feature. This process can be thought of as applying different levels of attention to each characteristic channel.

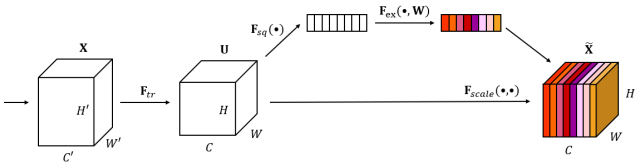

Fig. 1 Squeeze-and-Excitation Network (SENet)

In this way, each sample will have its own independent set of weights. In other words, the weights of any two samples are different. In SENet, the specific path to obtain the weight is shown in Fig. 2.

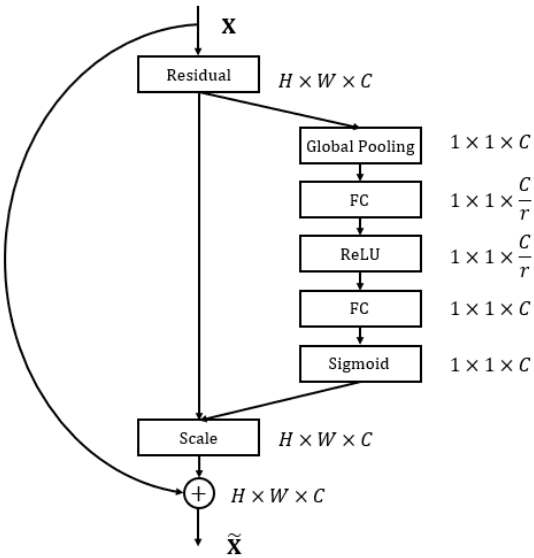

Fig. 2 The specific path to gain weight in SENet

### D. Soft Thresholding Under Deep Attention Mechanism

The deep residual shrinkage network draws lessons from the above-mentioned SENet sub-network structure to achieve soft thresholding under the deep attention mechanism. Through the sub-network in Fig. 3(a), a set of thresholds can be learned, and each characteristic channel can be soft-thresholded.

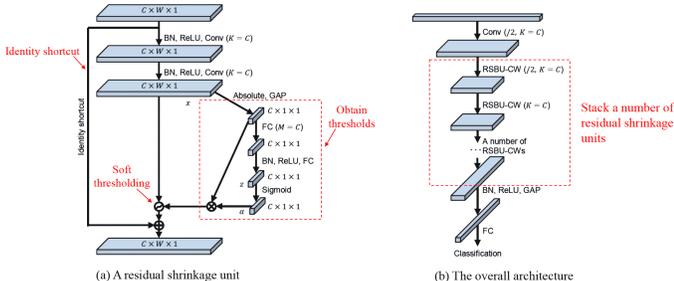

Fig. 3 (a) Building unit entitled RSBU. (b) Overall architecture of DRSN, where $K$ is the number of convolutional kernels in the convolutional layer, $M$ is the number of neurons in the FC network, and $C$, $W$, and 1 in $C \times W \times 1$ are the indicators of the number of channels, width, and height of the feature map, respectively. x, z, and α are the indicators of the feature maps to be used when determining thresholds [4].

In this sub-network, first find the absolute value of all the features of the input feature map. Then, after global mean pooling and averaging, a feature is obtained, denoted as A. In the other path, the feature map after global mean pooling is input into a small fully connected network. This fully connected network takes the Sigmoid function as the last layer, normalizes the output to between 0 and 1, and obtains a coefficient, denoted as α. The final threshold can be expressed as α×A. Therefore, the threshold is the average of a number between 0 and 1 the absolute value of the feature map. In this way, it not only ensures that the threshold is positive, but also not too large.

Moreover, different samples have different thresholds. Therefore, to a certain extent, it can be understood as a special attention mechanism: pay attention to features that are not related to the current task and set them to zero through soft thresholding; or, to pay attention to features related to the current task, keep them remain.

Finally, stack a certain number of basic modules and convolutional layers, batch normalization, activation functions, global mean pooling, and fully connected output layers to obtain a complete deep residual shrinkage network, as shown in Fig. 3(b).

### III. EXPERIMENTAL RESULTS

### A. Soft Thresholding Data Acquisition

The data set used for gesture classifications were downloaded from GitHub [8]. These data were recorded by MyoUP Armband with 8 channels. The database contains 69 sets of sEMG signals, and each set consists of separate recordings of 8 different gestures. For each gesture, the measurement lasts for one minute and it took a sampling frequency of 200Hz, so in total there are 200*60=12000 lines for one recording. The interval between a specific gesture motion and resting state was designed to be 5 seconds. Consequently, each recording has 10 samples of sEMG signals under the movement of a certain gesture and the whole dataset provides 10*8*69=5520 samples.

In this paper, all 8 different types of gestures were taken as targets to be classified, including hibernation, flexion, extension, radial deviation, ulnar deviation, pronation, supination and fist (Fig. 4). One special state among them is hibernation since it is considered as an approximative case of resting state. Various conditions of slight hand movements that are preferably ignored were uniformly grouped under this label.

Finally, the data was parceled into training and testing sets with 8 to 2 proportion. In order to give a clear evaluation of DRSN's performance against other mainstream machine learning models, classify results of identical sEMG signal by logistic regression model, random forest model and convolutional neural network (CNN) were respectively recorded for comparison. Furthermore, to corroborate the efficiency of DRSN's performance under high noise condition, another control group was arranged with the addition of noise to raw sEMG data. The processed data were then input into DRSN for training and testing.

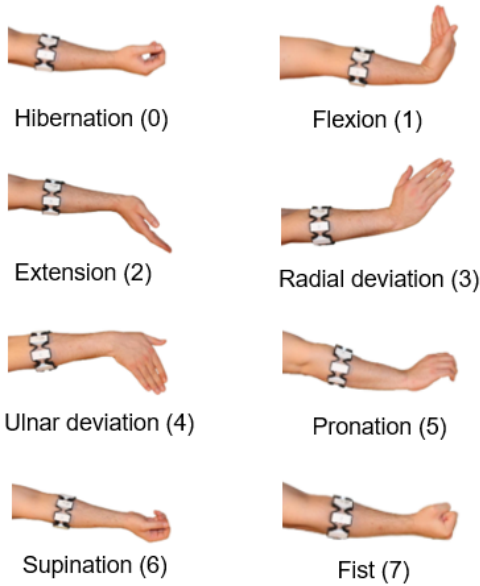

Fig. 4 Sketch map of eight types of gestures with their labels from 0-7 [8]

### B. Results

The training accuracy of DRSN and other three models are listed in TABLE I. Note that due to the limitations in equipment and time, only 9 samples out of 69 samples are used for training in this part. Among the four models, random forest attained the highest accuracy of 86%. The DRSN model reached a high accuracy of 81%, which could be higher after some further optimization in the structure and choosing of hyperparameters. In contrast, the results of logistic regression and CNN were unsatisfying. The poor performance of a linear model like logistic regression indicated that the group of EMG data does not have an inner linear relation. For CNN, the insufficiency in number of training samples was a possible reason for its low accuracy.

*TABLE I COMPARISON BETWEEN DRSN AND OTHER THREE MODELS*

| Model | Logistic regression | Random Forest | CNN | DRSN |
|---|---|---|---|---|
| Accuracy | 7% | 86% | 30% | 81% |

TABLE II gives the evaluation between the same DRSN model with different epochs. Compared to the case of 18 epochs, the accuracy of 31 epochs dropped for 3 percent, which may indicate that the classification curve became overfitted. However, these data are too inadequate to be a reference in the choosing of proper number of epochs. Further research requires more accuracy result of the DRSN model with different epochs to evaluate the tendency of accuracy when epoch number increases.

*TABLE II COMPARISON BETWEEN THE RESULTS OF DRSN WITH DIFFERENT EPOCHS*

| Epoch | 18 | 31 |
|---|---|---|
| Accuracy | 81% | 78% |

*TABLE III COMPARISON BETWEEN THE RESULTS OF NO NOISE/ADDED NOISE*

|  | No Noise | With Noise |
|---|---|---|
| Training Accuracy | 81%±16% | 75%±8% |
| Validation Accuracy | 80%±5% | 62%±3% |

Fig. 5, 6 gives the performance of DRSN under two different cases (no noise added, and noise added) while TABLE III shows the exact value. For the case of no noise added, the training accuracy kept a rapid growing trend in the first 20 epochs, while the rest epochs resulted in a frequent fluctuation rather than a stable increase. This may be a consequence of a large learning rate, so the network failed to reach the global minimum of loss function. The validation accuracy was close to training accuracy. In the training/validation loss graph, due to the large loss in first several epochs, the changing tendency of accuracy for the rest epochs were not clearly shown, which denoted that the setting of data points and unit length in axis need to be improved in further study. For the case of noise added, the training accuracy was lower compared to the no noise added case. It is also notable that there was a huge gap of approximately 20% between validation accuracy and testing accuracy, which illustrates that the current DRSN model has not been well trained for classification tasks under large noise.

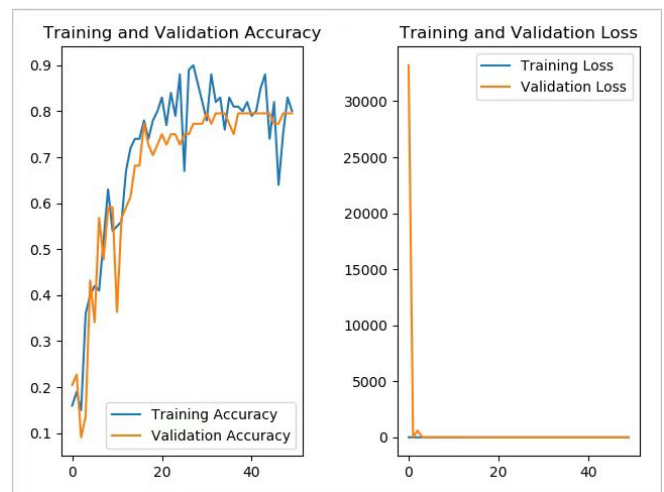

Fig. 5 Result of training and validation accuracy for DRSN, when no noise was added to the data

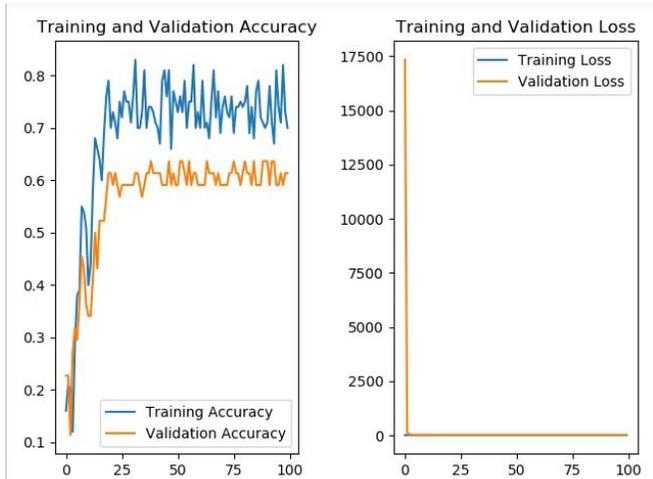

Fig. 6 Result of training and validation loss for DRSN, when no noise was added to the data

IV. CONCLUSIONS

The deep residual shrinkage network is a universal feature learning method. This is because in many features learning tasks, the samples will contain some noise and irrelevant information. These noises and irrelevant information may affect the effect of feature learning. for example:

In image classification, if the image contains many other objects at the same time, these objects can be understood as "noise"; the deep residual shrinkage network may be able to use the attention mechanism to notice these "noises", and then use soft Thresholding, setting the features corresponding to these "noises" to zero, it is possible to improve the accuracy of image classification.

In speech recognition, if the sound is relatively noisy, such as when chatting on the side of the road or in the factory workshop, the deep residual shrinkage network may improve the accuracy of speech recognition or give a way to improve the speech The idea of recognition accuracy.

In this paper, a deep residual shrinkage network is applied to achieve EMG based gesture identification. The network is tested using the Myo database downloaded from GitHub [8]. Experiments showed that DRSN model tend to become overfitted when the epoch exceed 30. DRSN performed well in classifying data without noise, but its accuracy of both testing and validation had an apparent drop when noise is added to raw EMG signal. When classifying the raw EMG data, DRSN showed an apparent advantage over logistic regression model and CNN model. However, its accuracy is slightly lower than random forest model. Generally, the DRSN model achieved a good classifying accuracy, but it had not reached the best state. One unneglectable factor is that these testing results were all under the condition that only 9 samples are used out of 69 samples. Additionally, the deciding of some hyperparameters can be further improved. In future research, it is expected that through necessary optimization in DRSN structure, it can have a better performance under high-noise condition.